\documentclass{optica-article}

\journal{opticajournal} 

\articletype{Research Article}

\usepackage{lineno}
\usepackage{enumitem}

\usepackage{xifthen}
\newcommand{\vq}[2][]{                
  \ifthenelse{\isempty{#1}}           %
    { \hat{\pmb{#2}} }                
    { \hat{\pmb{#2}}_\mathrm{#1} }    
}
\newcommand{\vqtil}[2][]{             
  \ifthenelse{\isempty{#1}}           %
    { \hat{\til{\pmb{#2}}} }                
    { \hat{\til{\pmb{#2}}}_\mathrm{#1} }    
}
\newcommand{\tq}[2][]{                
  \ifthenelse{\isempty{#1}}           %
    { \mathbb{#2} }                   
    { \mathbb{#2}_\mathrm{#1} }       
}
\newcommand{\vs}[2][]{                
  \ifthenelse{\isempty{#1}}           %
    { \mathbf{#2} }                   
    { \mathbf{#2}_\mathrm{#1} }       
}

\newcommand{\Tq}[3][]{                
  \ifthenelse{\isempty{#1}}           %
    { \mathbb{#3} }                   
    { \mathbb{#3}^{\rm #1}_{\rm #2} }       
}

\newcommand{\mr}[1]{\mathrm{#1}}

\begin{document}

\title{Proof-of-principle demonstration of a Polarization-Circulation Speed Meter}

\author{Yohei Nishino,\authormark{1,2*} Tomotada Akutsu,\authormark{2} Yoichi Aso,\authormark{2,3} Munetake Otsuka,\authormark{1,2} Luise Kranzhoff,\authormark{4,5} and Takayuki Tomaru\authormark{1,2,3,6,7}}

\address{\authormark{1}Department of Astronomy, University of Tokyo, Bunkyo, Tokyo 113-0033, Japan\\
\authormark{2}Gravitational Wave Science Project, National Astronomical Observatory of Japan (NAOJ), Mitaka City, Tokyo
181-8588, Japan\\
\authormark{3}The Graduate University for Advanced Studies (SOKENDAI), Mitaka City, Tokyo 181-8588, Japan\\
\authormark{4}Maastricht University, Department of Gravitational Waves and Fundamental Physics, 6200MD Maastricht, the Netherlands\\
\authormark{5}Nikhef, Science Park 105, 1098 XG Amsterdam, the Netherlands\\
\authormark{6}Accelerator Laboratory, High Energy Accelerator Research Organization (KEK), Tsukuba City, Ibaraki 305-
0801, Japan\\
\authormark{7}Institute for Cosmic Ray Research (ICRR), KAGRA Observatory, The University of Tokyo, Kashiwa City, Chiba 277-8582, Japan}

\email{\authormark{*}yohei.nishino@grad.nao.ac.jp} 


\begin{abstract*}
We present the first experimental implementation of a polarization-circulation speed meter. In our experiment, the interferometer was reduced to a single-cavity configuration with all mirrors fixed. A green-locking scheme was employed to stabilize the polarization circulation cavity, and a lock-acquisition procedure was demonstrated to realize speed-meter operation. The system was characterized by measuring the transfer function from a pseudo-displacement signal to the photodetector output, confirming that the device measures the speed of mirror motion.
These results support the feasibility of polarization-circulation speed meters and suggest that the control scheme could be extended to more complex configurations, such as Michelson interferometers and suspended-mirror systems.
\end{abstract*}

\section{Introduction}
A measurement’s quantum back-action is a fundamental phenomenon in precision metrology and has attracted considerable attention since the 1920s. According to Heisenberg’s uncertainty principle, when measuring the position of an object, the more precisely it is determined, the more strongly the canonically conjugate variable—momentum—is disturbed~\cite{1927ZPhy...43..172H}. Measurement back-action has been observed in a wide range of systems, including cold atoms, photonic crystal nanobeams, membranes, and macroscopic mirrors~\cite{murch2008observationquantummeasurementbackactionultracold, PhysRevLett.108.033602, PhysRevA.86.033840, Purdy_2013, Matsumoto_2015,Yu_2020}. Among these, gravitational-wave detectors have reached and even surpassed the standard quantum limit (SQL)~\cite{1968JETP.26.831, Yu_2020, Jia_2024}. The SQL is a direct consequence of the uncertainty principle, which characterizes the trade-off between measurement strength and backaction, providing a benchmark for quantum precision measurement.

\begin{figure}
    \centering
    \includegraphics[width=1.0\linewidth]{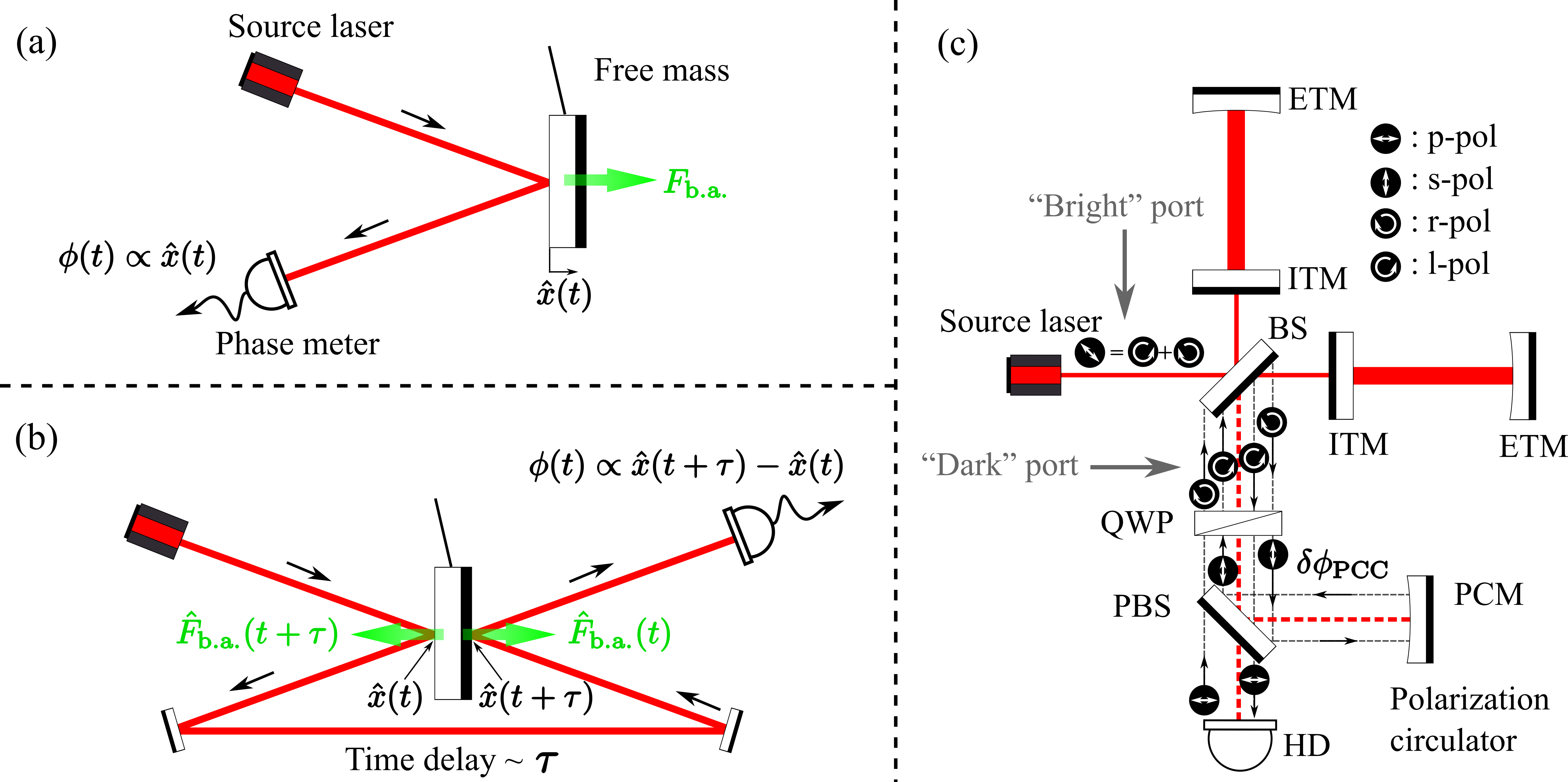}
    \caption{Toy models of the (a) position and (b) speed measurement. (c) Schematic of the polarization-circulation speed meter. For the direction of circulation, see the main text. Abbreviations: ITM, input test mass; ETM, end test mass; BS, beam splitter; QWP, quarter-wave plate; PBS, polarization beam splitter; PCM, polarization circulation mirror; HD, homodyne detector.}
    \label{fig:interferometer}
\end{figure}

The SQL arises because position operators at different times do not commute due to the back-action on momentum:
\begin{align}
    [\hat{x}(t), \hat{x}(t^\prime)] \neq 0.
\end{align}
This follows from the relation
\begin{align}
    \hat{x}(t^\prime) = \hat{x}(t) + \int_t^{t^\prime} \frac{\hat{p}(t^{''})}{m}\, \mathrm{d}t^{''}.
\end{align}
indicating that the back-action momentum or force ($F_\mathrm{b.a.}$, in Fig.~\ref{fig:interferometer}a) at the measurement time $t$ inevitably affect the uncertainty of the position in the future $t^\prime$.
By contrast, conserved quantities—such as momentum in the case of free masses—do commute at different times:
\begin{align}
    [\hat{p}(t), \hat{p}(t^\prime)] = 0.
\end{align}
Measurements of such observables are referred to as quantum non-demolition (QND) measurements~\cite{doi:10.1126/science.209.4456.547,1995qume.book.....B}. Since the kinetic momentum is proportional to velocity, momentum meters are often called \textit{speed meters}. Speed measurement can be realized by coherently measuring the mirror position twice with opposite signs, i.e., $x(t+\tau)-x(t)\simeq \bar{v}\tau$ (see Fig.~\ref{fig:interferometer}b), where $\tau$ is the delay between the first and second interactions. Various realizations of speed measurements for interferometric detectors have been proposed~\cite{PhysRevD.61.044002, BRAGINSKY1990251,PhysRevD.66.122004,Chen_2003,2018LSA.....7...11D, Knyazev_2018,nishino2025teleportationbasedspeedmeterprecision, bv8m-fpv9, kranzhoff2025demonstratingvelocityresponsetabletop}. Among these, the polarization-circulation speed meter proposed in Ref.~\cite{2018LSA.....7...11D} represents a highly feasible approach with the advantage that it preserves the conventional Michelson interferometer while requiring modification only at the detection port.

In this paper, we demonstrate a proof-of-principle experiment of the polarization-circulation speed meter. The speed meter employs two orthogonal polarizations that co-circulate in the Michelson interferometer and exert back-action forces in opposite directions, thereby canceling the measurement back-action with each other. Figure~\ref{fig:interferometer}c shows the layout of the scheme. The central part corresponds to a Fabry–P\'erot Michelson interferometer conventionally used for gravitational-wave detection, while three additional components are inserted at the dark port: a quarter-wave plate (QWP), a polarization beam splitter (PBS), and a polarization circulation mirror (PCM). The PBS selects only the p-polarized (p-pol) component of the incoming vacuum field, which is then converted to right-circular polarization (r-pol) by the QWP. The r-pol field couples with the classical laser amplitude entering from the bright port and exerts radiation-pressure back-action on the mirrors. The outgoing field is subsequently converted to s-pol, reflected by the PBS and PCM, and recycled back into the interferometer. This recycled field couples again with the classical amplitude and finally exits the dark port as p-pol. The classical field and the coherently recycled vacuum field are orthogonally polarized (r- and l-pol) and out of phase, such that the back-action forces acting on the mirrors exhibits opposite signs and cancels each other. 

To verify the principle of this method, our aims in the experiment are: (1) to build a system with polarization components that can measure the speed of mirror displacements, and (2) to demonstrate a control scheme for the speed meter, particularly the control of the polarization circulation mirror, based on the green-locking method proposed in Ref.~\cite{PhysRevD.107.084029}.
To this end, we simplified the interferometric layout to a single-cavity configuration (see Fig.~\ref{fig:Schematics}). 
In this experiment, the common and differential cavity modes were degenerate. 
Furthermore, since the observation frequency was higher than 10~kHz, the cavity mirrors were fixed. To simulate gravitational-waves, a phase-modulated beam was injected from the rear side of the end mirror. 
By measuring the system’s response from the applied phase modulation to the output signal, we observed speed-meter–type behavior as expected from the theory in Sec.~\ref{sec:Theory}. 
The control scheme tested in this setup can be scaled to more complex systems, such as Michelson interferometers and suspended-mirror configurations.

The remainder of this paper is organized as follows. 
In Sec.~\ref{sec:Theory}, we provide a brief theoretical analysis of the speed-meter response, including imperfections. 
In Sec.~\ref{sec:Layout}, the experimental setup including the lock acquisition method are described. 
In Sec.~\ref{sec:Results}, the results of the speed measurement are reported. 
Finally, we discuss about the results and conclude in Sec.~\ref{sec:Conclusion}.

\section{Theory}\label{sec:Theory}

In this section, we present a theoretical analysis of the speed-meter response. The results are used to fit the measured transfer function in Sec.~\ref{sec:Results}, Fig~\ref{fig:TF_with_inset}.
When the carrier field resonate in a single optical cavity, the reflectivity as a function of the sideband frequency $r(\Omega)$ can be written in the first-order approximation as
\begin{align}
    r(\Omega) &\simeq \frac{\gamma_1-\gamma_2+i\Omega}{\gamma_1+\gamma_2-i\Omega},
\end{align}
where $\Omega$ is the frequency of an audio sideband generally caused by mirror motion. The parameters $\gamma_1$ and $\gamma_2$, which represent the bandwidths associated with the input coupling rate of laser light into the cavity via the ITM and the loss, are defined as
\begin{align}
    \gamma_1 &\equiv \frac{cT_\mathrm{ITM}}{4l_\mathrm{cav}}, \quad
    \gamma_2 \equiv \frac{c\mathcal{L}_\mathrm{cav}}{4l_\mathrm{cav}}, \label{eq:gamma_2}
\end{align}
where $T_\mathrm{ITM}$ is the power transmissivity of the input test mass (ITM), $\mathcal{L}_\mathrm{cav}$ is the round-trip loss of the main cavity, and $l_\mathrm{cav}$ is the cavity length. Note that the delay $\tau$ corresponds to the cavity strage time, i.e., $\tau=2\pi/\gamma_1$.

\begin{figure}
    \centering
    \includegraphics[width=0.5\linewidth]{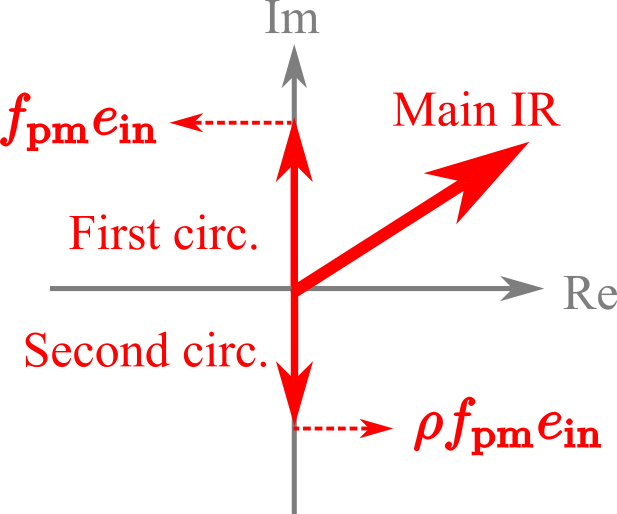}
    \caption{Phasor diagram. The modulated input sidebands, shown in dashed arrow were denoted as $e_\mathrm{in}$. The first circulation field had a transfer function of $f_\mathrm{pm}$, and the second circulation field had an additional factor $\rho$. The measured transfer function was obtained by taking the beat note between the sideband and the main IR field.}
    \label{fig:phaser}
\end{figure}

The input and output modulated sidebands are denoted as $e_\mathrm{in,out}(\Omega)$ respectively  which are proportional to the voltages $V_\mathrm{in,out}$ (see Fig.~\ref{fig:Schematics} and Fig.~\ref{fig:evaluation_schemes}b). The transfer function of the speed meter, $f_\mathrm{sm}(\Omega)$, is proportional to the subtraction of the first and second circulations and to the response function of the position meter, $f_\mathrm{pm} \propto 1/(\gamma_1-i\Omega)$:
\begin{align}
    f_\mr{sm}(\Omega) = \frac{V_\mathrm{out}(\Omega)}{V_\mathrm{in}(\Omega)} &\propto \frac{1-\rho(\Omega)}{2}f_\mr{pm} \notag\\
    &\propto \frac{1-e^{i(\delta\phi_{\mathrm{PCC}}+\delta\phi^\mr{IR}_\mr{ret})}(1-\mathcal{L}_\mathrm{PCC})r(\Omega)}{2}\, f_\mr{pm} \notag\\
    &\simeq \frac{\gamma_2+\mathcal{L}_\mathrm{PCC}\gamma_1/2-i\{\gamma_1(\delta\phi_\mr{PCC}+\delta\phi^\mr{IR}_\mr{ret})/2+\Omega\}}{\gamma_1-i\Omega}\, f_\mr{pm}, \label{eq:3}
\end{align}
where $\delta\phi_\mathrm{PCC}$ is the root-mean-square (RMS) phase fluctuation of the PCC, $\delta\phi^\mr{IR}_\mr{ret}$ is the retardation error of the QWP, and $\mathcal{L}_\mathrm{PCC}$ is the round-trip optical power loss inside the PCC. 
The second circulation has a factor $\rho(\Omega) = e^{i(\delta\phi_{\mathrm{PCC}}+\delta\phi^\mr{IR}_\mr{ret})}(1-\mathcal{L}_\mathrm{PCC})r(\Omega)$ as shown in Fig.~\ref{fig:phaser}.
The PCC length was controlled such that 
\begin{align}
    \phi_\mathrm{PCC} = 2\pi\frac{l_\mathrm{PCC}}{\lambda_0} \equiv \pi \; (\mathrm{mod}\; 2\pi).
\end{align}
The main IR field serves as a local oscillator, linearly projecting $f_\mr{sm}(\Omega)$ onto the output voltage.
We use the approximations $\cos\theta\simeq1$ and $\sin\theta\simeq\theta$, and neglect higher-order terms in $\mathcal{L}_\mr{cav}$, $\mathcal{L}_\mr{PCC}$, and $\delta\phi_\mr{PCC}$.

The effective loss of the PCC can be expressed as
\begin{align}
    \mathcal{L}_\mathrm{PCC} 
    = 2 \left(\mathcal{L}_\mathrm{QWP} + T_\mathrm{SPBS} + R_\mathrm{PPBS}\right) 
    + T_\mathrm{PCM} + \mathcal{L}_\mathrm{align} + \mathcal{L}_\mathrm{mis}, 
    \label{eq:31}
\end{align}
where $\mathcal{L}_\mathrm{QWP}$, $\mathcal{L}_\mathrm{align}$, and $\mathcal{L}_\mathrm{mis}$ denote the losses in the QWP, due to misalignment, and due to mode mismatching, respectively. 
$T_\mathrm{SPBS}$ is the transmissivity of the PBS for s-polarization, $R_\mathrm{PPBS}$ is the reflectivity of the PBS for p-polarization, and $T_\mathrm{PCM}$ is the transmissivity of the PCM.
The cutoff frequency is therefore written as
\begin{align}
    \gamma_\mathrm{cut} &= \gamma_2+\frac{\mathcal{L}_\mathrm{PCC}\gamma_1}{2} \\
    &= \frac{c}{4l_\mathrm{cav}}\left(\mathcal{L}_\mathrm{cav}+\frac{\pi\mathcal{L}_\mathrm{PCC}}{\mathcal{F}}\right).
\end{align}

We measure the response in the phase (or sine) quadrature of the outgoing field. 
Therefore, the response function is expressed in the quadrature representation:
\begin{align}
    \psi(\Omega)=\frac{f(\Omega)-f^*(-\Omega)}{2i}.
\end{align}
Our observable is defined as the ratio between the speed and position transfer functions to cancel out the calibration uncertainty:
\begin{align}
    H(f) \equiv \frac{\psi_\mr{sm}(2 \pi f)}{\psi_\mr{pm}(2 \pi f)} = \frac{\gamma_\mr{cut}/2 \pi-i f}{\gamma_1/2 \pi-i f},
\end{align}
which is used in Sec.~\ref{sec:Results} (see also Ref.~\cite{PhysRevD.107.084029}).

\section{Experiment setup}\label{sec:Layout}
\subsection{Layout}
\begin{figure}
    \centering
    \includegraphics[width=1.0\linewidth]{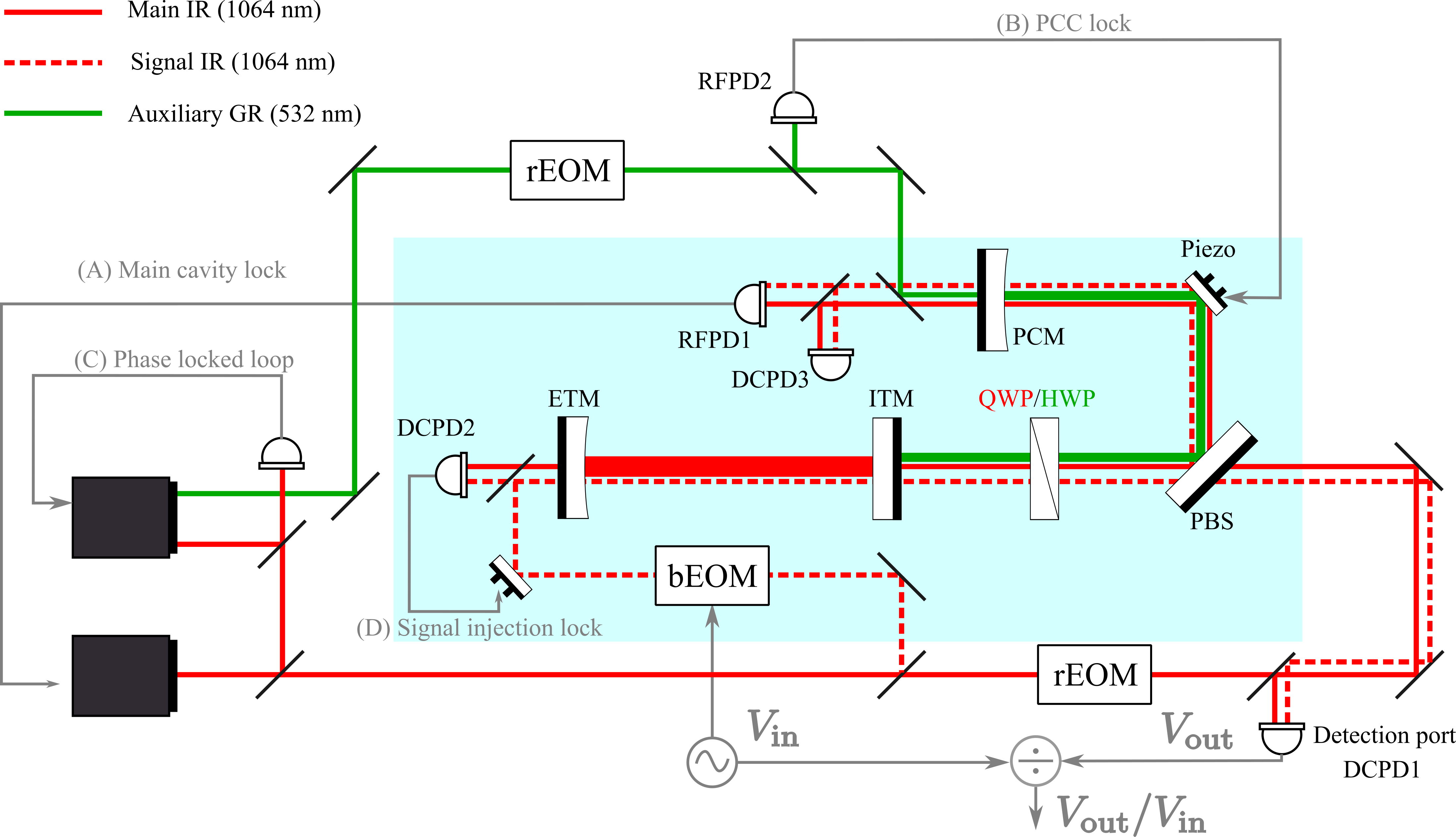}
    \caption{Schematic of the experimental setup. Red lines represent the 1064~nm laser beam. The solid red line indicates the main beam used to lock the main cavity, while the dashed red line indicates the signal beam that carried the phase signal. The main beam is injected through the polarization beam splitter (PBS), and the PDH signal is derived from the transmission of the polarization circulation mirror (PCM). The signal beam is picked off from the main beam and injected from the anti-reflective side of the end test mass (ETM) after being modulated by the broadband electro-optic modulator (bEOM). The green (GR) beam is injected from the anti-reflective side of the PCM and resonated inside the polarization circulation cavity (PCC). Gray lines represent electrical signals. The cyan shaded area is where we zoom into in Fig.~\ref{fig:evaluation_schemes}.}
    \label{fig:Schematics}
\end{figure}

Figure~\ref{fig:Schematics} shows a schematic of the speed-meter setup. The main cavity is formed by an input test mass (ITM) and an end test mass (ETM). The polarization circulator consists of a wave plate, a polarization beam splitter (PBS), and a polarization circulation mirror (PCM). A high-reflectivity mirror mounted on a piezo actuator is used to control the length of the polarization circulation cavity (PCC). The wave plate functions as a quarter-wave plate (QWP) at 1064~nm and as a half-wave plate (HWP) at 532~nm. 

\begin{figure}
    \centering
    \includegraphics[width=1.0\linewidth]{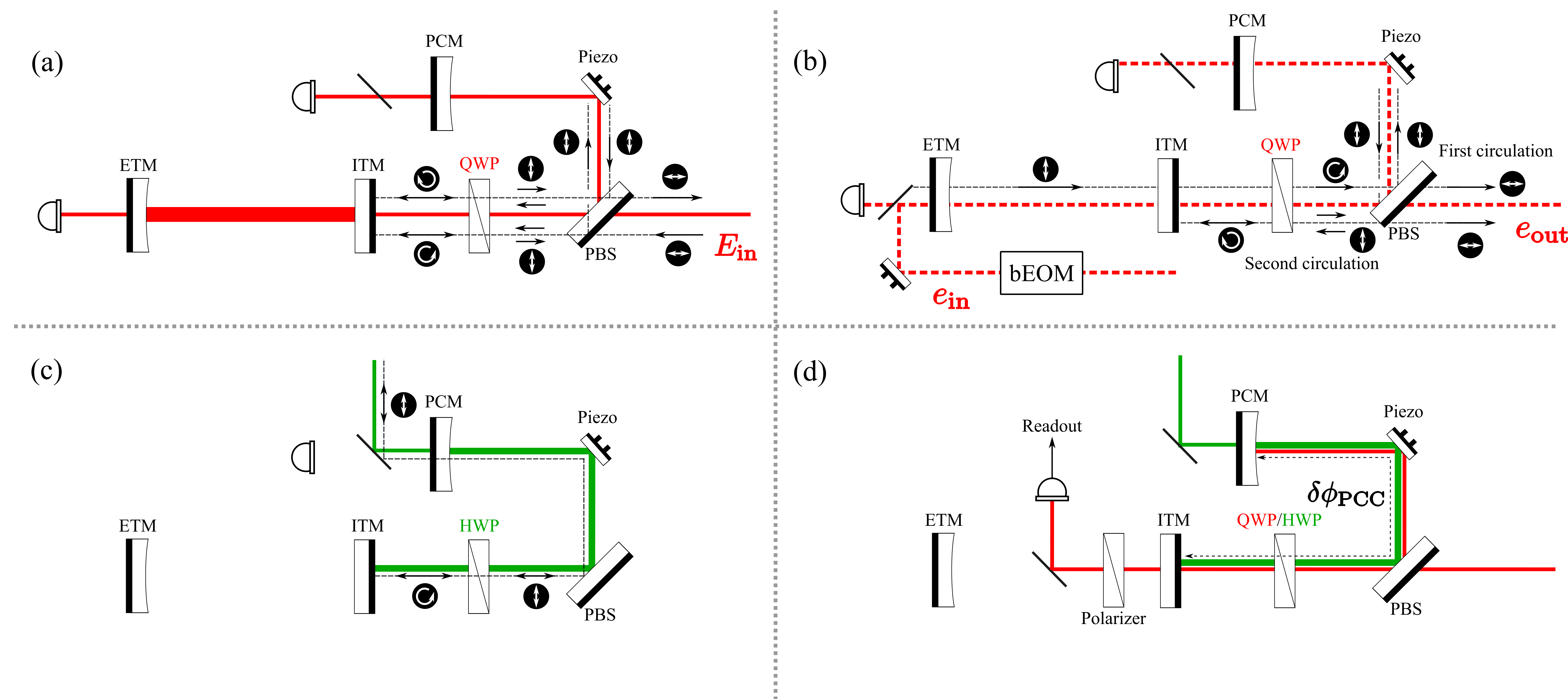}
\caption{(a) Path of the main IR beam. The input state was p-pol. The photodetector behind the ETM monitored the transverse mode leaking from the main cavity. Approximately 1\% of the light transmitted through the PCM and was used for frequency locking of the main cavity via the PDH method. 
(b) Path of the signal IR beam. The input, denoted as $e_\mathrm{in}$, was demodulated by the broadband EOM. The QWP converted the input s-pol to r-pol, and the PBS transmitted half of the light (first circulation). The remaining component, which was s-pol, was recycled back into the cavity and transmitted through the PBS at the end. The interference between the first and second circulations was detected as $e_\mathrm{out}$. 
(c) Path of the GR beam. The wave plate functioned as a half-wave plate (HWP). The input s-pol beam was converted to l-pol by the HWP and then converted back to s-pol. In this way, the GR beam resonated inside the PCC, allowing the PDH method to be employed. The ITM transmissivity at 532~nm was chosen to be sufficiently low so that the GR beam did not simultaneously resonate in the main cavity. 
(d) Setup for evaluating the out-of-loop noise. The PCC was locked by the GR beam. A polarizer placed inside the cavity selected a linear polarization, forming a Michelson-type interferometer in the polarization domain. By taking the interference between the circulating IR beams, the setup measured the path fluctuation of the PCC.}
    \label{fig:evaluation_schemes}
\end{figure}

We show the paths of the three beams used in the experiment in Fig.~\ref{fig:evaluation_schemes}a-c. The main infrared (IR, 1064~nm) beam (depicted as a solid red line, see Fig.~\ref{fig:evaluation_schemes}a) enters the main cavity through the PBS. The beam is modulated at 15~MHz by a resonant electro-optic modulator (rEOM) to generate a control signal for the main cavity using the Pound–Drever–Hall (PDH) method. The PDH signal is obtained by detecting a small transmission through the PCM (at RFPD1) and is fed back to the frequency of the main laser. A portion of the main beam is picked off and used as the signal beam (dashed red line, see Fig.~\ref{fig:evaluation_schemes}b). This signal beam is passed through a broadband electro-optic modulator (bEOM), which introduced signal sidebands, and is phase-locked to the main beam transmitted through the cavity.

The green (GR, 532~nm) beam (solid green line, see Fig.~\ref{fig:evaluation_schemes}c) is generated by a separate laser source that is phase-locked to the main laser. This beam is injected into the PCC after being modulated at 20~MHz by another rEOM located at the rear side of the PCM. Since the wave plate functions as a half-wave plate at 532~nm, the PCC forms a low-finesse cavity ($\mathcal{F}_\mathrm{GR}\sim 50$) for the GR beam, thereby enabling the use of PDH locking. The resulting PDH signal obtained through RFPD2 is fed back to the piezo actuator to stabilize the PCC length. Parameters at 532~nm are listed in Tab.~\ref{tab:GR-parameters}.

The transfer function is measured at a DC photodetector (DCPD1) located after the rEOM (referred to hereafter as the ``detection port''). At this port, the reflected main beam (serving as the local oscillator) and the transmitted signal beam (serving as the signal) interferes to produce a beat note. The system ultimately measures the transfer function from the voltage applied to the bEOM to the output voltage of the DCPD1.

\begin{table}[htbp]
\caption{Parameters of the infrared (IR) control used in the experiment. Values are nominal unless otherwise indicated by (m) for measured.}
  \label{tab:IR-parameters}
  \centering
\begin{tabular}{ccc}
\hline
Parameter & Value & Note  \\ \hline\hline
    $\lambda_0$ [nm] & 1064 & Nd:YAG fundamental wavelength \\ 
    $T_\mathrm{ITM}$ & 4000 ppm & ITM transmissivity \\ 
    $T_\mathrm{ETM}$ & 35 ppm & ETM transmissivity \\ 
    $T_\mathrm{PCM}$ & 1\% & PCM transmissivity (m) \\ 
    $\delta\phi^\mathrm{IR}_\mathrm{ret}/2\pi$ & $7(\pm1)\times10^{-3}$ & QWP retardation error \\ 
    $\mathcal{L}_\mathrm{PCC}$ & 3\% & Total losses in PCC (m) \\ 
    $l_\mathrm{cav}$ [m] & 0.15 & Main cavity length (m) \\ 
    $l_\mathrm{PCC}$ [m] & 0.38 & Mean PCC length (m) \\ \hline
    $\mathcal{F}$ & 1500 & Finesse of the main cavity \\ 
    $f_c$ [Hz] & $3.2\times10^5$ & Cavity pole frequency\\ 
\hline
\end{tabular}
\end{table}

\begin{table}[htbp]
\caption{Parameters of the green (GR) beam. }
  \label{tab:GR-parameters}
  \centering
\begin{tabular}{ccc}
\hline
Parameter & Value & Note  \\ \hline\hline
    $\lambda_\mathrm{GR}$ [nm] & 532 & GR laser wavelength \\ 
    $T_\mathrm{PCM}^\mathrm{GR}$ & 1\% & PCM transmissivity (m) \\ 
    $T_\mathrm{ITM}^\mathrm{GR}$ & 500 ppm & ITM transmissivity \\ 
    $\mathcal{F}_\mathrm{GR}$ & $\sim 50$ & Finesse of PCC \\ 
\hline
\end{tabular}
\end{table}

\subsection{Control degrees of freedom}

The system has four control degrees of freedom:
\begin{enumerate}[label=(\Alph*)]
    \item The frequency of the main laser with respect to the length of the main cavity 
    \item The PCC length with respect to the frequency of the green (GR) beam, 
    \item the relative frequency between the main and auxiliary laser sources, and
    \item the relative phase between the signal and main beams.
\end{enumerate}
(A) and (B) are controlled using the Pound–Drever–Hall (PDH) method. Resonant electro-optic modulators (rEOMs, shown in Fig.~\ref{fig:Schematics}) is employed for phase modulation at 15~MHz and 20~MHz, respectively.  

The loop corresponding to (C) is a phase-locked loop (PLL), whose control signal is obtained by demodulating the beat note between the two lasers. The frequency difference could be tuned arbitrarily by adjusting the frequency of the local oscillator (LO). Owing to this tunability, the PCC length could be adjusted relative to the frequency of the main IR (see phase~(iv) in Sec.~\ref{ssec.LA}).

Finally, (D) is controlled by detecting the beat note between the transmission of the main IR from the main cavity and the reflection of the signal beam from the ETM. The two optical fields are locked at mid-fringe to generate an error signal that is linear with respect to the phase difference.

\subsection{Lock acquisition scheme}\label{ssec.LA}
\begin{figure}
    \centering
    \includegraphics[width=1.0\linewidth]{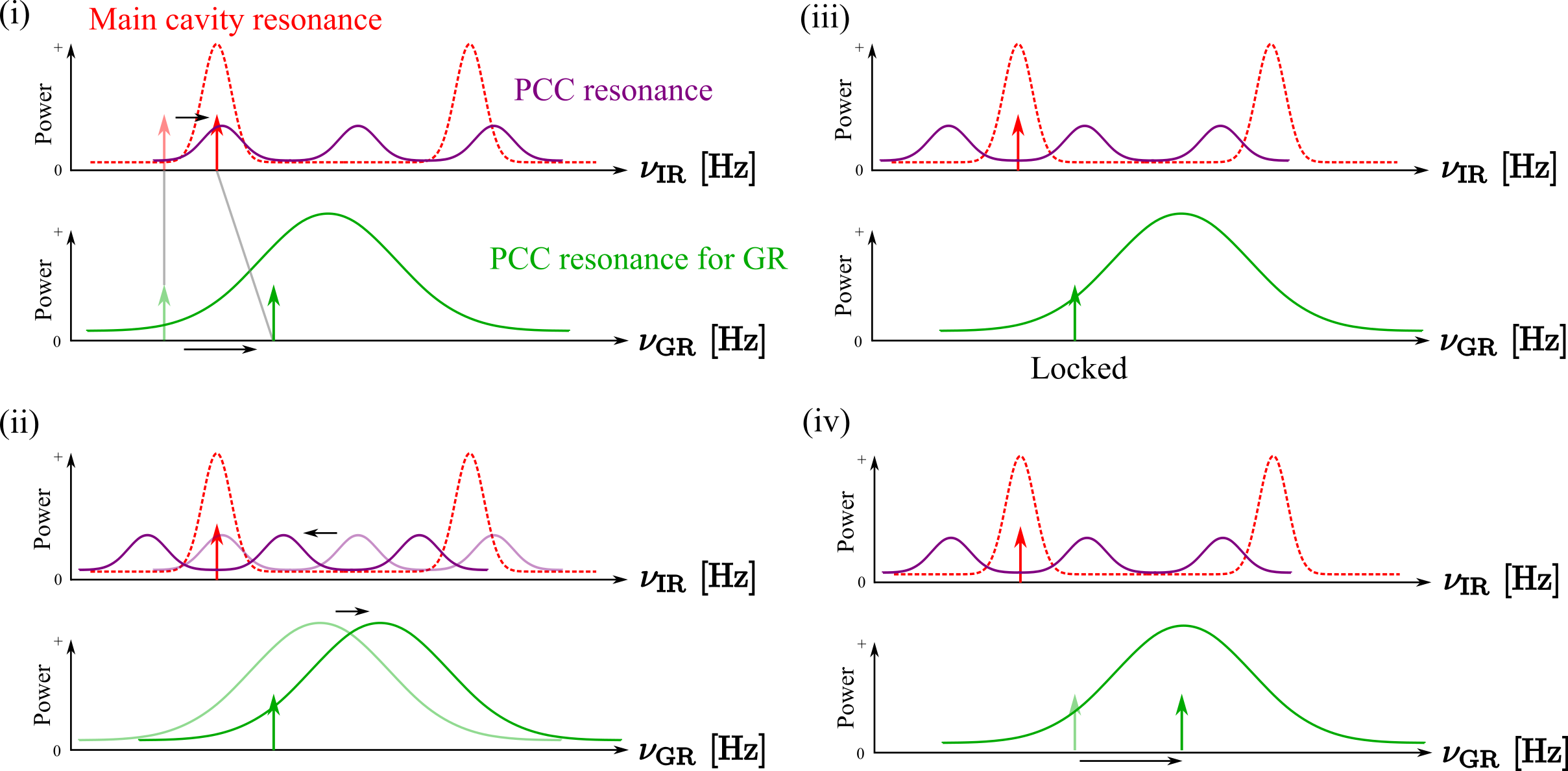}
    \caption{Schematic of lock acquisition. Each of the four panels shows the stored power in the cavities as a function of light frequency. The dashed red and solid purple curves indicate the resonances of the main cavity and the PCC for the main IR beam, respectively. The solid green curve shows the resonance of the PCC for the GR beam. In each panel, the translucent shapes indicate the states prior to each procedure.}
    \label{fig:LockAcqflow}
\end{figure}

To operate the system as a speed meter, we first need to draw the system into the optimal working point, a process called lock acquisition. Fig.~\ref{fig:LockAcqflow} shows the schematic of the acquisition. The procedure is as follows:

\begin{enumerate}[label=(\roman*)]
  \item The frequency of the main laser is locked to the length of the main cavity.
  \item The PCC length is scanned near the locking point for the GR beam. 
  \item The PCC length is locked to the GR frequency.
  \item The frequency of the local oscillator (LO) in the PLL is scanned to bring the PCC length to the optimal point for the IR. 
\end{enumerate}
After fixing the PLL LO frequency, the PCC length is locked with respect to the IR frequency. After this step, the system operates as a speed meter.
Note that the PLL should already be locked before this procedure.

\section{Results}\label{sec:Results}
In this section, we present two key results of the experiment: the lock acquisition of the system (Fig.~\ref{fig:lockAcquisition}) and the measurement of the classical transfer function of the speed meter (Fig.~\ref{fig:TF_with_inset}). 
Here, ``classical'' refers to the case where the input to the system is an artificial phase modulation of the light (coherent state), rather than quantum fluctuations. 
The measured transfer function was fitted with the theoretical prediction, as described in Sec.~\ref{sec:Theory}. 
In addition, we demonstrate that the stability of the PCC is sufficient to maintain the speed-meter operating condition.

\subsection{Lock acquisition}
\begin{figure}
    \centering
    \includegraphics[width=1.0\linewidth]{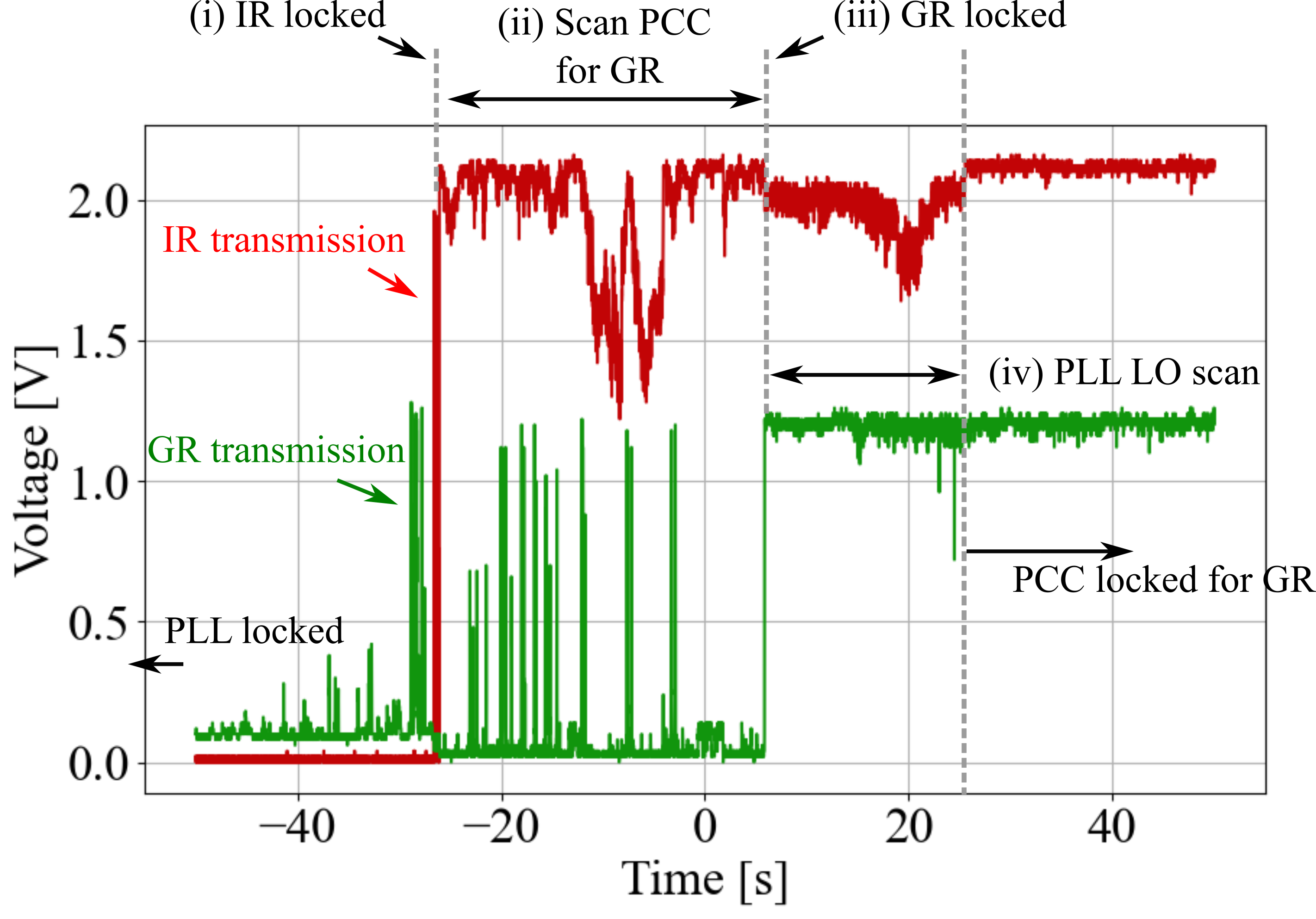}
    \caption{Time-series data during lock acquisition. The red and green traces represent the IR transmission power from the main cavity and the GR transmission power from the PCC, respectively. The frequencies of the main IR and GR lasers were phase-locked before this measurement. The locking procedure proceeded as follows: (i) the main IR frequency was locked to the length of the main cavity, (ii) the PCC length was scanned to search for the GR locking point, (iii) the PCC length was locked to the GR frequency, and (iv) the LO frequency of the PLL was scanned to tune the PCC to the optimal point for the IR. After that, the PCC length was finally optimized for the system to function as a speed meter.}
    \label{fig:lockAcquisition}
\end{figure}

The time-series data obtained during the acquisition phase are shown in Fig.~\ref{fig:lockAcquisition}. The optimal PCM position corresponds to the maximum transmission of the IR beam at DCPD2 (red trace). Under this condition, the interference at DCPD1 was destructive, and the system operated as a speed meter. 

The decrease in GR power observed upon IR lock at step (i) was due to the residual IR component, not fully removed by the dichroic mirror, being transmitted through the main cavity. 
In step (ii), the cavity length was scanned by varying the voltage applied to the piezo actuator on one mirror of the PCC, in order to bring the PCM position close to the locking point, during which the GR light flashed. 
During procedure (iv), the GR transmission power remained constant because the PCC length followed the GR frequency, whereas the IR transmission power fluctuated, reflecting the motion of the PCC.

After achieving full lock, we proceeded with the measurement of the transfer function and the evaluation of the PCC length stability, as described below.

\subsection{Transfer function}
\begin{figure}
    \centering
    \includegraphics[width=1.0\linewidth]{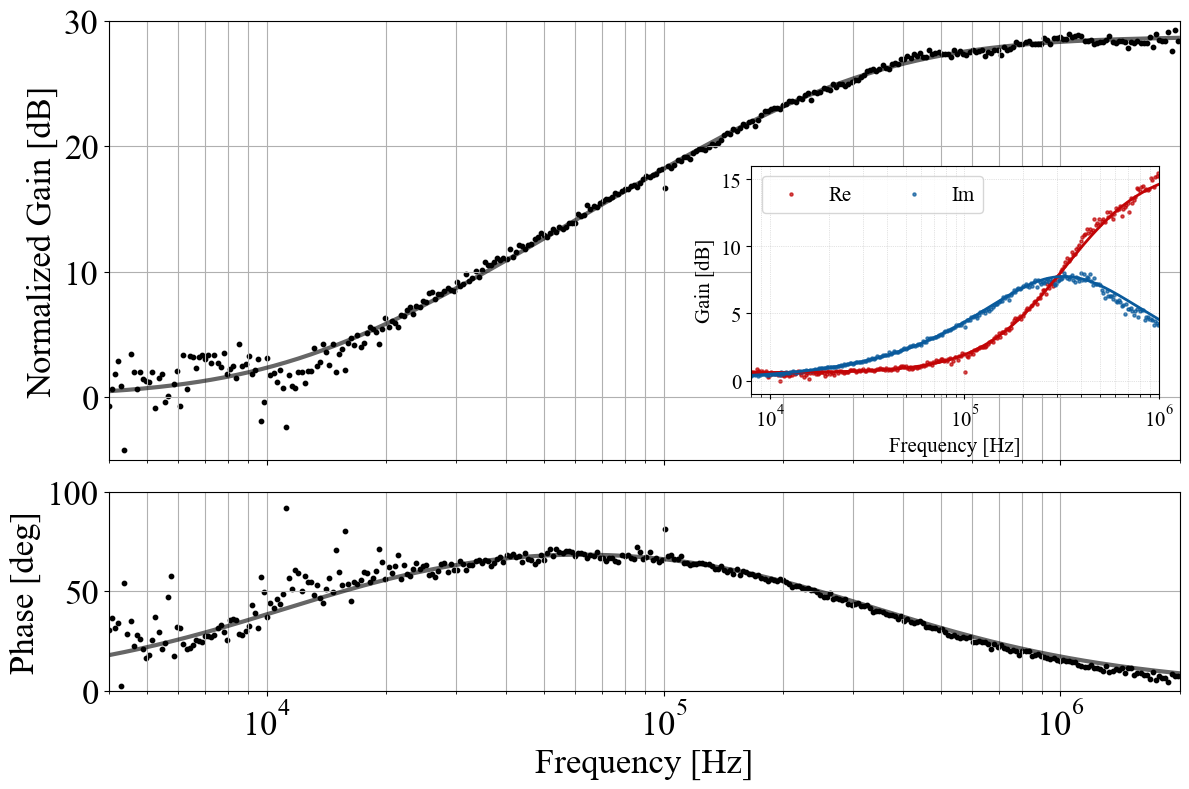}
    \caption{Transfer function of the speed meter in a Bode diagram. Circles denote the measured data, and solid curves indicate the fitted results. The inset shows the decomposition into the real (Re) and imaginary (Im) parts used for fitting. Error bars are not shown; see the main text for discussion in Sec.~\ref{sec:Results}.}
    \label{fig:TF_with_inset}
\end{figure}

Figure~\ref{fig:TF_with_inset} shows the measured transfer function across a frequency range from 4~kHz to 2~MHz. The gain of the transfer function was normalized to its DC value of the fitting curve. The transfer function was further normalized by the position-meter transfer function, which was measured simultaneously with a DCPD3 placed at the rear side of the PCM (see Fig.~\ref{fig:Schematics}).

The loss in the main cavity ($\mathcal{L}_\mathrm{cav}$) was treated as a free parameter:
\begin{equation}
    H_\mathrm{fit}(2\pi f; g, \mathcal{L}_\mathrm{cav}) = g\, H(2\pi f; \mathcal{L}_\mathrm{cav}), \label{eq:fit1}
\end{equation}
where $g$ represented an overall gain factor, which is calibrated using the values around 2~MHz. The cost function to be minimized was defined as
\begin{align}
    \Xi &= |H_\mathrm{m}(f) - H_\mathrm{fit}(f; g, \mathcal{L}_\mathrm{cav})|^2 \notag\\
    &= \mathrm{Re}\big[H_\mathrm{m}(f) - H_\mathrm{fit}(f; g, \mathcal{L}_\mathrm{cav})\big]^2 
     + \mathrm{Im}\big[H_\mathrm{m}(f) - H_\mathrm{fit}(f; g, \mathcal{L}_\mathrm{cav})\big]^2,
\end{align}
where $H_\mathrm{m}(f)$ denoted the measured data. The decomposition into the real and imaginary parts is shown in the inset of Fig.~\ref{fig:TF_with_inset}. The fitting yielded an estimated value of $\mathcal{L}_\mathrm{cav} \sim 85$~ppm. 

Most importantly, the characteristic frequency dependence of the speed meter---namely, that the transfer function increases proportionally to frequency---was clearly observed. 
The natural cutoff frequency was $\sim 3.2\times10^5$~Hz, corresponding to the pole of the main cavity. 
The low-frequency cutoff indicates that the speed measurement was degraded by imperfections. 
In the quantum regime, this cutoff frequency corresponds to the point below which the back-action is not canceled and only the mirror position can be measured.

Error bars were not available for the transfer function because the spectrum analyzer provided only averaged values, which did not allow estimation of standard deviations. Despite this limitation, the statistical independence of the data points and the close agreement between the fitted curves and the experimental results justify treating the outcomes as reliable for the purposes of the present demonstration.

\subsection{PCC stability}
\begin{figure}
    \centering
    \includegraphics[width=0.95\linewidth]{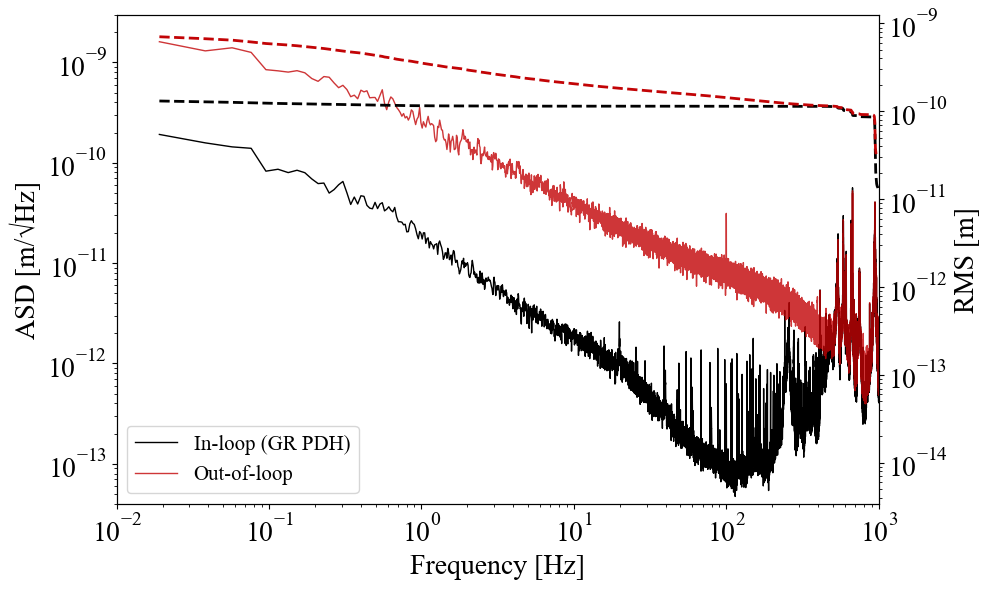}
    \caption{Amplitude spectral density (ASD) of the PCC length (solid). Results from the in-loop and out-of-loop measurements are shown in black and red, respectively. The accumulated root-mean-square (RMS) noise is plotted as dashed curves, with the corresponding scale shown on the right axis.}
    \label{fig:ASD_PLL_locked_budget_RMS}
\end{figure}

We evaluated the stability of the PCC length using both in-loop and out-of-loop measurements (see Fig.~\ref{fig:ASD_PLL_locked_budget_RMS}). 
The in-loop noise was obtained by projecting the error signal in the main PDH loop onto the PCC length, whereas the out-of-loop noise corresponded to the length fluctuation directly measured with the IR beam. The out-of-loop evaluation setup is shown in Fig.~\ref{fig:evaluation_schemes}d. A polarizer was placed inside the main cavity. Because of this polarizer, the light transmitted through the ITM interfered with the circulating light recycled by the ITM, PBS, and PCM. Consequently, the setup measured the PCC fluctuation in the same manner as a Michelson interferometer.
The two results were consistent at high frequencies ($>400$~Hz) but diverged at lower frequencies. This discrepancy is not fully understood, but it may have originated from nontrivial noise in the control loop or from residual IR polarization converted into intensity fluctuations in the out-of-loop measurement.
The total noise was quantified in terms of the RMS value below $\sim20$~mHz, yielding an RMS of $1\times10^{-10}$~m for the out-of-loop measurement and $7\times10^{-10}$~m for the in-loop measurement. 

\section{Discussion and Conclusion}\label{sec:Conclusion}
From Eq.~\eqref{eq:3}, the effects of imperfections can be estimated. 
Using the parameters listed in Tab.~\ref{tab:IR-parameters}, the cutoff frequency was $\gamma_\mathrm{cut}/2\pi \sim 6.8$~kHz, and the detuning due to retardation error was $7.1$~kHz, which produced the cutoff observed in Fig.~\ref{fig:TF_with_inset}. 
The detuning due to PCC fluctuations can be derived from the formula $\gamma_\mathrm{PCC} = \gamma_1 \delta\phi_\mathrm{PCC}/2$ (see Sec.~\ref{sec:Theory}). 
Using the RMS value obtained from the out-of-loop measurement, the detuning was estimated as $\gamma_\mathrm{PCC}/2\pi \approx 700$~Hz, affecting frequencies well below the cutoff. 
We therefore conclude that the green-locking method was adequate for the purposes of this experiment.

At relatively low frequencies ($<20$~kHz) in Fig.~\ref{fig:TF_with_inset}, the data points deviated. 
This behavior arose from the reduced signal-to-noise ratio at lower frequencies and from the measurement approaching the unity-gain frequency of the main IR loop ($\sim 5$~kHz), where signal injection destabilized the system. 
This was acceptable in the present experiment, since the observation range was set below the control band. 
However, for practical use in gravitational-wave detectors targeting the frequency range of $1$--$100$~Hz, calibration of the signal within the control band will be required. We leave this point for future work.

We have performed a proof-of-principle demonstration of a polarization-circulation speed meter. In this work, the gravitational-wave detector was approximated by a single-cavity system with all mirrors fixed. Using the green-locking scheme, the PCC length was controlled to enable the system to operate as a speed meter. By injecting a pseudo-displacement signal and measuring the transfer function at the photodetector output, we confirmed that the system exhibited speed-meter–type behavior with $\propto f$ dependence in the transfer function. It is expected that this approach could be applied in the future to more complex systems, such as Michelson interferometers or suspended-mirror configurations.

\section*{Acknowledgments}
The authors are grateful to Kentaro Komori for valuable discussions regarding the experimental setup. 
LK acknowledges support from the European Research Council (Advanced grant 101019978).
This research is supported by JSPS Grant-in-Aid for JSPS Fellows Grant Number 23KJ0787 and 23K25901.  We also thank the Advanced Technology Center (ATC) of NAOJ for providing the experimental space and technical support.

\bibliography{ref}

\end{document}